\documentclass[12pt]{article}
\usepackage[cp1251]{inputenc}
\usepackage[english]{babel}
\usepackage{amssymb,amsmath}
\usepackage{graphicx}

\title{ Possible measurement of {\it P}-states probability in the ground state $^4$He nucleus }

\author{Yu.P.Lyakhno}

\begin {document}

\maketitle

\begin{center} {\it National Science Center "Kharkov Institute of
Physics and Technology" \\ 61108, Kharkiv, Ukraine}
\end{center}

\begin{abstract}

Using the experimental data on total {\it S}=1 transitions
cross-sections for $^4$He$(\gamma,p)$$^3$H and
$^4$He$(\gamma,n)$$^3$He reactions as the base, the paper discussed
the possibility of measuring the probability of $^3P_0$-states in
the ground state $^4$He nucleus. The analysis of the experimental
data has suggested the conclusion, that within the statistical
error, the ratio of the cross section of the reaction in the
collinear geometry to the cross section of the electrical dipole
transition with the spin {\it S}=0 at the angle of nucleons emission
$\theta_N$=$90^{\circ}$ $\nu_p$ and $\nu_n$ in the range of photon
energies 22$\le$E$_{\gamma}$$\le$100 MeV doesn't depend from the
photon energy. This is in agreement with the assumption that the
{\it S}=1 transitions can originate from $^3P_0$ states of the
$^4$He nucleus. Average values of magnitude $\nu_p$ and $\nu_n$ in
the mentioned photon energy range are calculated
$\nu_p$=0.01$\pm$0.002 and $\nu_n$=0.015$\pm$0.003. The errors are
statistical only.

\vskip20pt


\end{abstract}

\section{Introduction}

The model-independent calculation of the ground state of the
nucleus, and also of its scattering states, can be carried out on
the basis of realistic inter-nucleonic forces and exact methods of
solving the many-nucleon problem. The $^4$He nucleus can serve as a
good test for setting this approach to work. In \cite {1} have
calculated the ground states of the lightest nuclei using the
realistic  {\it NN} Argonne AV18 \cite {2} and CD Bonn \cite {3}
potentials, and also, the 3{\it N} forces UrbanaIX \cite{4} and
Tucson-Melbourne \cite{5,6}. The calculations were carried out using
the Faddeev-Yakubovsky (FY) technique \cite {7,8}, which was
generalized by Gloeckle and Kamada (GK) \cite {9} to the case of
taking into account two- and three-nucleon forces.  The authors have
estimated the error of nuclear binding energy calculations for
$^4$He to be $\sim$50 keV. The calculated binding energy appeared to
be $\sim$200 keV higher than the experimentally measured value. In
view of this, the authors drew the conclusion that there is a
possible contribution of the 4{\it N} forces that could have a
repulsive character. Another possible explanation of this result
might be the inconsistency of the data on {\it NN} and 3{\it N}
forces.

The tensor part of the {\it NN} interaction and the 3{\it NF's}
forces generate the $^4$He nuclear states with nonzero orbital
momenta of nucleons. Table 1 gives the probabilities of $^1S_0$,
$^3P_0$ and $^5D_0$ states of the $^4$He nucleus calculated in
\cite{1} (notation: $^{2S+1}L_J$). The calculations gave the
probability of $^5D_0$ states having the total spin {\it S}=2 and
the total orbital momentum of nucleons {\it L}=2 of the $^4$He
nucleus to be $\sim16\%$, and the probability of $^3P_0$ states
having {\it S}=1 and  {\it L}=1 to be 0.75\%. It is obvious from
Table 1 that the consideration of the 3{\it NF's} contribution
increases the probability of $^3P_0$ states by a factor of $\sim$2.

 {\bf T a b l e 1. The $^1S_0$, $^3P_0$, and $^5D_0$ states probabilities for
the ground state $^4$He nucleus (in percentage terms).}
\begin{center}
\begin{tabular}[t]{|c|c|c|c|}
\hline Interaction
& $^1S_0,\%$ &  $^3P_0,\%$ & $^5D_0,\%$  \\
\hline
AV18 & 85.87 & 0.35 & 13.78 \\
CD-Bonn & 89.06 & 0.22 & 10.72 \\
AV18+UIX & 83.23 & 0.75 & 16.03 \\
CD-Bonn+TM & 89.65 & 0.45 & 9.9  \\
\hline
\end{tabular}
\end{center}

In \cite{10}, Kievsky {\it et al.} have calculated the ground states
of the lightest nuclei by the method of hyperspherical harmonic,
using  the {\it NN} and 3{\it N} potentials calculated from the
effective field theory. Various variants of the mentioned potentials
predict the contribution from $^3P_0$ states of the $^4$He nucleus
to be between 0.1\% and 0.7\%. Thus, the measurement of the
probability of states with nonzero orbital momenta of nucleons can
provide a new information about inter-nucleonic forces.

\section{The analisys of the experimental data about cross-section
$^4$He$(\gamma,p)$$^3$H and $^4$He$(\gamma,n)$$^3$He reactions in
the collinear geometry}

Here, we discuss the possibility of measuring the probability of
$P(^3P_0)$ states of the ground state $^4$He nucleus through the
studies of two-body ($\gamma,p$) and ($\gamma,n$) reactions of
$^4$He. In these reactions, transitions matrix elements of two
types, with spins {\it S}=0 and {\it S}=1 of the final state of the
particle system, may take place. It is known \cite{11}, that at the
electromagnetic interaction the spin-flip of hadronic particle
system is significantly suppressed. The  {\it S}=1 transitions may
originate from $^3P_0$ nuclear states with no spin-flip. Maybe such
transitions can occur also from $^1S_0$  or $^5D_0$ states of the
$^4$He nucleus as a result of different channels of the reaction
coupled, for example, with the existence of states with nonzero
orbital momentums of nucleons of the residual nucleus and from the
secondary effects. It can be supposed that the cross section of the
($\gamma$,N)  reaction doesn't depend on the total spin of the
ground state of $^4$He nucleus. Then the ratio
\begin{equation}
\label {eq1} \alpha=\frac{\sigma(^3M_{1,2})}{\sigma_{tot}(\gamma,N)}
\end{equation}
of the total cross sections of the transitions with the spin {\it
S}=1 to the total cross section $\sigma_{tot}(\gamma,N)$ of the
reaction, after the subtraction of the contribution of other
possible mechanisms of formation of transitions with spin {\it S}=1,
can be sensible to contribution of {\it P}-wave component in the
wave function of $^4$He nucleus. The indexes (1,2) are total
momentums 1$^-$,1$^+$ and 2$^+$ of final-state of particle system at
the transitions {\it S}=1.

In the {\it E1}, {\it E2} and {\it M1} approximation, the laws of
conservation of the total momentum and parity for two-body
$(\gamma,p)$ and $(\gamma,n)$ reactions of $^4$He nuclear
disintegration permit the occurrence of two multipole transitions
$E1^1P_1$ and $E2^1D_2$ with the spin {\it S}=0 and four transitions
$E1^3P_1$, $M1^3D_1$, $M1^3S_1$ and $E2^3D_2$ with the spin {\it
S}=1 of final-state particles. According to the present experimental
data the sum of total cross sections of transitions with the spin
{\it S}=1 is $\sim10^{-2}$ of the total cross section of the
reaction. The nucleon emission distributions in the polar angle for
each of the mentioned transitions are presented in Table 2.

{\bf T a b l e 2: Angular distributions for {\it E1}, {\it E2} and
{\it M1} multipoles.}
\begin{center}
\begin{tabular}[t]{|c|c|c|}
\hline Spin of the & Multipole &    Angular \\ final-states &
transition & distribution  \\ \hline S=0  & $|E1^1{\rm P}_1|^2$ &
$\sin^2\theta$
\\ & $|E2^1{\rm D}_2|^2$ & $\sin^2\theta$$\cos^2\theta$\\ \hline &
$|E1^3{\rm P}_1|^2$ & 1+$\cos^2\theta$\\   S=1 & $|M1^3{\rm S}_1|^2$
& const \\  & $|M1^3{\rm D}_1|^2$ & 5-3$\cos^2\theta$\\ & $|E2^3{\rm
D}_2|^2$ &
1-3$\cos^2\theta$+4$\cos^4\theta$\\
\hline
\end{tabular}
\end{center}

It can be seen from Table 2 that the reaction cross-section in the
collinear geometry can be due only to {\it S}=1 transitions, at that
d$\sigma(0^{\circ}$)=d$\sigma(180^{\circ})$. For the purpose of
determining the reaction cross-section in the collinear geometry, an
analysis was made of the information available in the literature
about differential cross sections of the $^4$He$(\gamma,p)$$^3$H and
$^4$He$(\gamma,n)$$^3$He reactions in the photons energy range up to
the meson-producing threshold.

In \cite{12,13}, the reaction products were registered at
nucleon-exit polar angles $0^0\le\theta_N\le180^0$, using chambers
placed in the magnetic field. However, the number of events
registered in those experiments was insufficient for measuring the
reaction cross-section in the collinear geometry (the cross-section
estimation is shown by full circles in Fig.1).

Jones {\it et al.} \cite{14} have measured the differential cross
section for the  $^4$He($\gamma,p$)$^3$H reaction at tagged photon
energies between 63 and 71 MeV (triangles in Figs. 1 and 2). The
reaction products were registered by means of a wide-acceptance
detector LASA. The measurements were performed in the interval of
polar proton-exit angles $22.5^0\le\theta_p\le 145.5^0$. The lack of
data for large and small angles of proton escape has led to
significant errors in the measurement of the reaction cross-section
in the collinear geometry.

In \cite{15}, (cross in Fig.1), a monoenergetic photon beam in the
energy range from 21.8 to 29.8 MeV and nearly a 4$\pi$ time
projection chamber were used to measure the total and differential
cross-sections for photodisintegration reactions of $^4$He nucleus.
The authors found that the M1 strength was about 2$\pm$1\% of the E1
strength.

The differential cross sections for two-body ($\gamma,p$) and
($\gamma,n$) reactions have been measured by Arkatov {\it et al.}
\cite{16}, \cite{17} in the bremsstrahlung photon energy range from
the reaction threshold up to $E_{\gamma}$=150 MeV. The reaction
products were registered with the help of a diffusion chamber,
placed in the magnetic field in the interval of polar nucleon-exit
angles $0^0\le\theta_N\le180^0$. Later on, Nagorny {\it et al.}
\cite{18} reprocessed this experiment, using a new program for
geometric remodeling of events, a more powerful (for that time)
computer, and also, an upgraded particle track measuring system. The
number of the processed events was increased by a factor of 3, and
amounted to $\sim3\cdot10^4$ for each of the ($\gamma$,p) and
($\gamma$,n) reaction channels. The differential cross-sections were
measured with a 1 MeV step up to $E_{\gamma}$=45 MeV, and with a
greater step at higher energies, as well as with a 10$^{\circ}$
c.m.s. step in the polar nucleon-exit angle. The authors have
published their data on the differential cross-sections at photon
energies of 22.5, 27.5, 33.5, 40.5, 45, and 49 MeV. The
comprehensive data of Arkatov {\it et al.} on the differential
cross-sections for the $^4$He$(\gamma,p)$$^3$H and
$^4$He$(\gamma,n)$$^3$He reactions can be found in ref. \cite{19}.

The differential cross-section for this reactions in the c.m.s. can
be presented as:
\begin{equation}
\label {eq2} \frac{\rm d\sigma }{\rm d\Omega }=
A[{\sin^2\theta(1+\beta\cos\theta
+\gamma\cos^2\theta)+\varepsilon\cos\theta+\nu}],
\end{equation}
where
$\nu$=[d$\sigma(0^{\circ}$)+d$\sigma(180^{\circ})$]/2d$\sigma_1(90^{\circ})$,
and
$\varepsilon$=[d$\sigma(0^{\circ}$)-d$\sigma(180^{\circ})$]/2d$\sigma_1(90^{\circ})$,
here d$\sigma_1(90^{\circ})$ is a cross section $E1^1P_1$ transition
at the nucleon emission angle $\theta_N=90^0$.

It can be supposed, that transition $M1^3{\rm S}_1$ is the main one
only at the reaction threshold \cite{20}. In the majority of works,
it was supposed that the $E2^3{\rm D}_2$ amplitude is the smallest
one. This suggestion is confirmed by experimental hints \cite{21}.
Assuming that basic transitions, which give a contribution to ratio
$\nu$, are electric dipole transitions with spin {\it S}=1 and {\it
S}=0 and executing integration of proper angular distributions over
the solid angle, we obtain:

\begin{equation}
\label {eq3} \alpha=\frac{\sigma(E1^3{\rm P}_1)}{\sigma(E1^1{\rm
P}_1)}=\frac{\rm d\sigma(0^0) }{\rm d\sigma_1(90^0)}=\nu
\end{equation}

If it is supposed, that the main transition with the spin {\it S}=1
is the $M1^3{\rm D}_1$ transition, then

\begin{equation}
\label {eq4} \alpha=\frac{\sigma(M1^3{\rm D}_1)}{\sigma(E1^1{\rm
P}_1)}=\frac{3\rm d\sigma(0^0)}{\rm d\sigma_1(90^0)}=3\nu
\end{equation}

The ratio $\nu$ was calculated as a result of least-squares fitting
(LSM) of expression (2) to the experimental data on differential
cross-sections \cite{19} (with a double step in the photon energy).
The results of calculations are presented in Figs. 1 and 2 as open
circles. It can be seen from the figures that within the statistical
errors the ratio of the cross-section in collinear geometry of the
{\it S}=1 transitions to the cross-sections at polar nucleon-exit
angle $\theta_N$=$90^{\circ}$ reaction in the photon energy region
22$\le$E$_{\gamma}$$\le$100 MeV is independent of the photon energy.
This is in agreement with the assumption that these transitions
might originate from {\it P}-states of the $^4$He nucleus. The
average values of the ratio $\nu$ in the mentioned photon energy
range are calculated to be $\nu_p$=0.019$\pm$0.002 and
$\nu_n$=0.028$\pm$0.003. The average values of the coefficients are
$\varepsilon_p$=0.$\pm$0.002 and $\varepsilon_n$=-0.001$\pm$0.003.

\begin{figure}[h]
\noindent\centering{
\includegraphics[width=90mm]{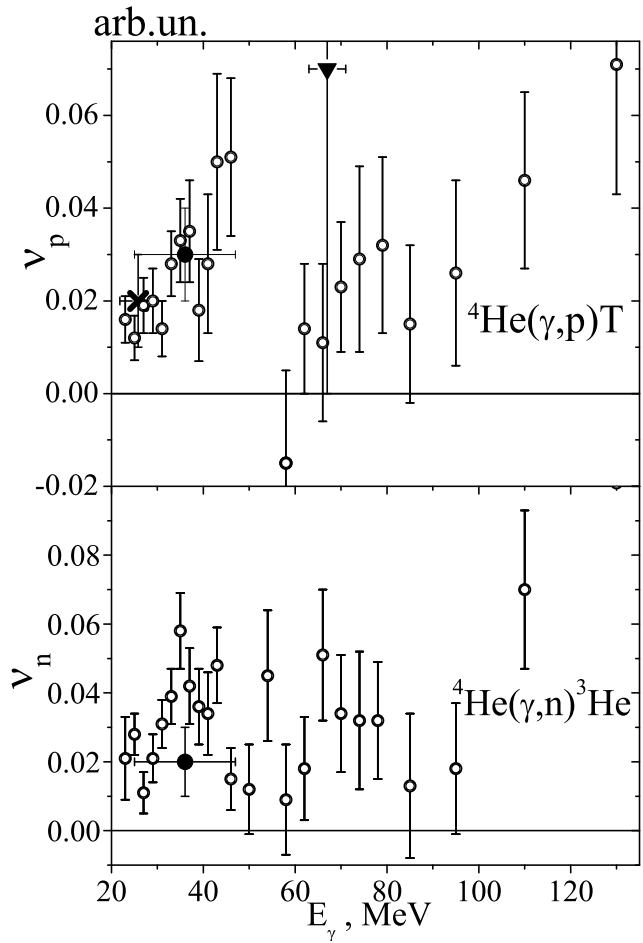}
} \caption{Coefficients $\nu_p$ and $\nu_n$. The cross shows the
data from \cite{15}; triangle - data from \cite{14}; full circles -
data from \cite{13}; open circles-data from \cite{19}. The errors
are statistical only.}
\end{figure}

\begin{figure}[h]
\noindent\centering{
\includegraphics[width=90mm]{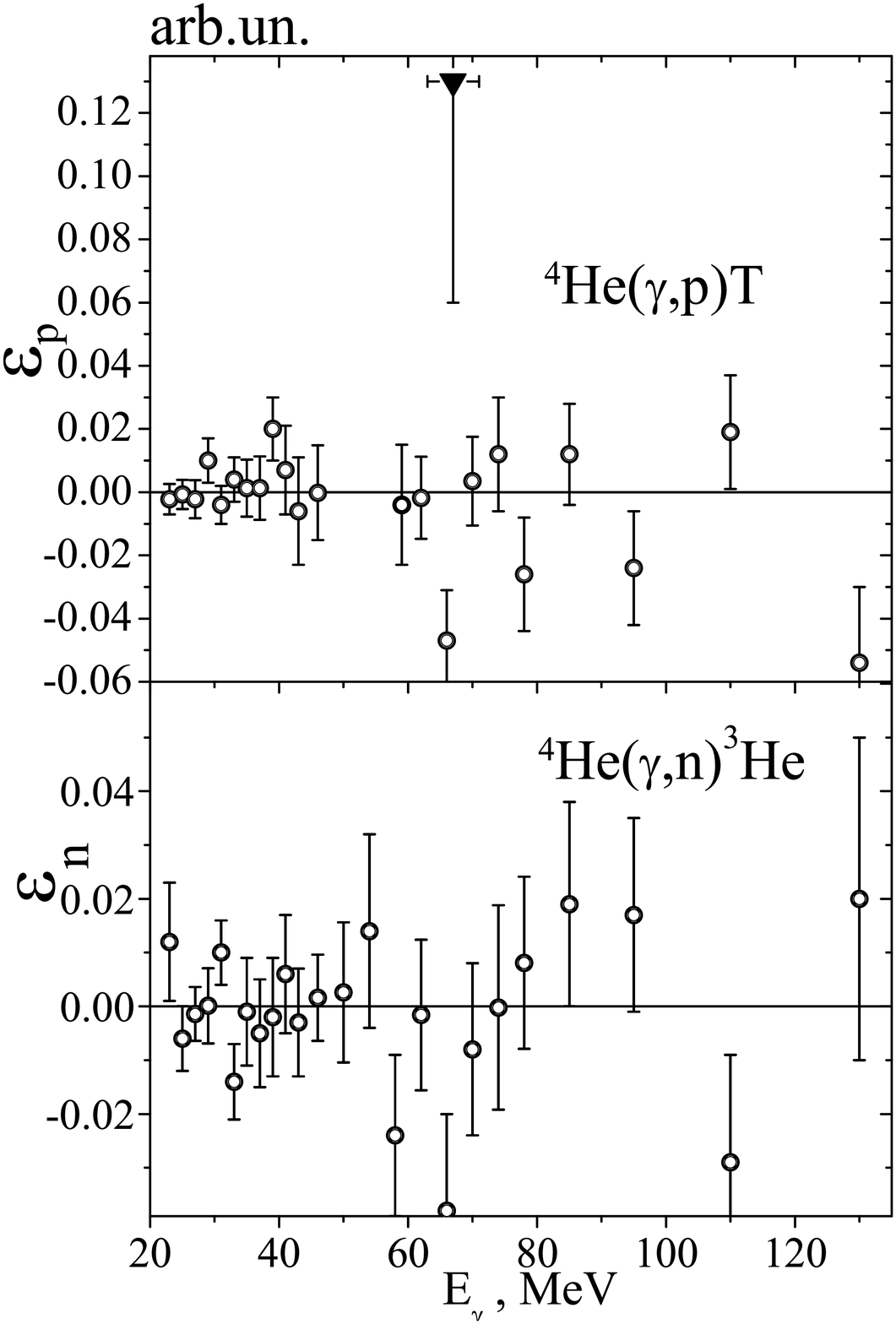}
} \caption{Coefficients $\varepsilon_p$ and $\varepsilon_n$.
Triangle-data from \cite{14}; Open circles-data from \cite{19}. The
errors are statistical only.}
\end{figure}

The calculated $\nu_p$ and $\nu_n$  values may be displaced as a
result of experimental data histogramming, and also, due to polar
nucleon-exit angle measurement errors, which were
$\delta\theta_N$=0.5$^{\circ}$$\div$1$^{\circ}$, as reported in ref.
\cite{22}. In this connection, by the use of simulation we have
determined  the corrections for the histogramming step as
10$^{\circ}$  and for $\delta\theta_N$=1$^{\circ}$ \cite{23}. Taking
into account these corrections, we have $\nu_p$=0.01$\pm$0.002 and
$\nu_n$=0.015$\pm$0.003.

The systematic error of the data might be caused by the inaccuracy
in the measurement of the resolution on the polar angle of the
nucleon emission. In particular, the difference in coefficients
$\nu_p$ and $\nu_n$ might be conditioned by the fact that resolution
of the neutron $\delta\theta_n$ emission angle was worse than of the
angle of the proton $\delta\theta_p$ emission. Besides, a small
number of events in some histogramming steps, especially at high
photon energies, can lead to a systematic error specified by the use
of the LSM method. The conclusion in \cite{24} about large errors in
the cross-section measurements in the collinear geometry had been
based on early works of Arkatov {\it et al.} \cite{16}, \cite{17}.

The cross-sections of spin {\it S}=1 transitions can be measured by
means of polarization observables. For example, the transitions
$E1^1P_1$ and $E2^1D_2$ with the spin {\it S}=0  exhibit the
asymmetry of the cross-section with linearly polarized photons
$\Sigma(\theta)$=1 at all polar nucleon-exit angles, except
$\theta_N$=$0^{\circ}$ and $180^{\circ}$. The difference of the
asymmetry $\Sigma$ from unity can be due to spin {\it S}=1
transitions. With an aim of separating the contributions from
$E1^3P_1$, $M1^3D_1$ and  $M1^3S_1$ transitions, Lyakhno {\it et
al.} \cite{21} performed a combined analysis of both the
experimental data on the cross-section asymmetry $\Sigma(\theta)$
and the data on differential cross-sections for the reactions under
discussion \cite{19} at photon energies $E_{\gamma}^{peak}$= 40, 56
and 78 MeV. Because of small cross-sections of spin {\it S}=1
transitions, the errors of asymmetry $\Sigma(\theta)$ measurements
have led to considerable errors in the cross-section measurements of
these transitions.

The authors of \cite{20,25,26} investigated the reactions of
radiative capture of polarized protons by tritium nuclei. In
\cite{25}, Wagenaar {\it et al.} have investigated the capture
reaction at proton energies 0.8$\le E_p\le$9 MeV. They came to the
conclusion that the main transition with spin {\it S}=1 is
$M1^3S_1$. In \cite{20}, this reaction was investigated by Pitts at
the proton energy E$_p$=2 MeV.  It was concluded that the main
transition with {\it S}=1 is $E1^3P_1$. These contradictory
statements were caused by considerable statistical and systematic
errors of the experimental data. Within the experimental errors, the
data obtained in the studies of ($\gamma,N$) and ($\vec p,\gamma$)
reactions are in satisfactory agreement between themselves
\cite{21}.

\section{Conclusions}

In \cite{20} it was found that the transitions with spin {\it S}=1
can be conditioned by the contribution of meson exchange currents
(MEC). It should be noticed, that the MEC contribution depends on
the photon energy \cite{27}. Despite the considerable MEC
contribution into the total cross section of the reaction,
contribution of the spin-flip of the hadronic particle system can be
insignificant. The weak dependence of the ratio of {\it S}=1 to {\it
S}=0 transitions cross-sections from the photon energy in the energy
region from the reaction threshold up to E$_{\gamma}\sim$100 MeV
(this corresponding to the nucleon momentum P$_N$$\sim$350 MeV/c)
may point to an insignificant contribution to the total
cross-section of spin {\it S}=1 transitions by the final-state
particle interactions, and also, by other photon energy-dependent
reaction mechanisms. The present experimental data coincide with the
supposition that contribution $^3P_0$ components of the ground state
of $^4$He nucleus to the formation of the transitions with the spin
{\it S}=1 at the two-body ($\gamma$,N) reaction can be considerable
and these data can be used for measurement of contribution {\it
P}-wave component in the wave function $^4$He nucleus.

A number of investigations, e.g. \cite{28}-\cite{32}, were made into
the reaction $^2$H($\vec d,\gamma)^4$He with an aim of measuring the
probability of $^5D_0$ states of the $^4$He nucleus. This reaction
permits the occurrence of three types of transitions with {\it S}=0,
1 and 2. In this connection, the analysis of the experimental data
on this reaction may be more complicated than that of the two-body
($\gamma,N$) reaction. It might be reasonable to perform a combined
analysis of these reactions. The detailed theoretical calculations
of this reactions are required.

The author is thankful to Prof. R.~Pywell, Drs. L.~Levchuk and
A.~Buki for the discussion of the article; thanks are also due to a
group of people of the Kharkiv Institute of Physics and Technology
for their experimental data on differential cross-sections for
reactions $^4$He$(\gamma,p)$$^3$H and $^4$He$(\gamma,n)$$^3$He.

\end{document}